\documentclass[]{aa}
\usepackage{graphicx}
\usepackage{setspace}
\usepackage{natbib}
\usepackage{longtable}
\usepackage{amssymb}
\usepackage[varg]{txfonts}
\bibpunct{(}{)}{;}{a}{}{,}

\begin{document}

\newcommand{\FeII}{[\ion{Fe}{ii}]}
\newcommand{\TiII}{[\ion{Ti}{ii}]}
\newcommand{\SII}{[\ion{S}{ii}]}
\newcommand{\OI}{[\ion{O}{i}]}
\newcommand{\OIp}{\ion{O}{i}}
\newcommand{\PII}{[\ion{P}{ii}]}
\newcommand{\NI}{[\ion{N}{i}]}
\newcommand{\NII}{[\ion{N}{ii}]}
\newcommand{\NIp}{\ion{N}{i}}
\newcommand{\NiII}{[\ion{Ni}{ii}]}
\newcommand{\CaIIp}{\ion{Ca}{ii}}
\newcommand{\PI}{[\ion{P}{i}]}
\newcommand{\CIp}{\ion{C}{i}}
\newcommand{\HeI}{\ion{He}{i}}
\newcommand{\MgIp}{\ion{Mg}{i}}
\newcommand{\MgIIp}{\ion{Mg}{ii}}
\newcommand{\NaI}{\ion{Na}{i}}
\newcommand{\HI}{\ion{H}{i}}
\newcommand{\brg}{Br$\gamma$}
\newcommand{\pab}{Pa$\beta$}

\newcommand{\macc}{$\dot{M}_{acc}$}
\newcommand{\lacc}{L$_{acc}$}
\newcommand{\lbol}{L$_{bol}$}
\newcommand{\mjet}{$\dot{M}_{jet}$}
\newcommand{\mh}{$\dot{M}_{H_2}$}
\newcommand{\Ne}{n$_e$}
\newcommand{\h}{H$_2$}
\newcommand{\kms}{km\,s$^{-1}$}
\newcommand{\um}{$\mu$m}
\newcommand{\lam}{$\lambda$}
\newcommand{\msyr}{M$_{\odot}$\,yr$^{-1}$}
\newcommand{\Av}{A$_V$}
\newcommand{\msun}{M$_{\odot}$}
\newcommand{\lsun}{L$_{\odot}$}
\newcommand{\cm}{cm$^{-3}$}
\newcommand{\ergscm}{erg\,s$^{-1}$\,cm$^{-2}$}

\newcommand{\bet}{$\beta$}
\newcommand{\alfa}{$\alpha$}

\hyphenation{mo-le-cu-lar pre-vious e-vi-den-ce di-ffe-rent pa-ra-me-ters ex-ten-ding a-vai-la-ble excited in-ter-fe-ro-me-ters em-pi-ri-cal ca-li-bra-tion vi-si-bi-li-ty vi-si-bi-li-ties con-ti-nuum a-na-ly-sis 
cha-rac-te-ris-tics tem-pe-ra-tu-re co-rres-pon-ding in-ter-fe-ro-me-tric ro-ta-ting mo-dels mag-ne-tos-phe-re}

\title{Probing the accretion-ejection connection with VLTI/AMBER  }
\subtitle{High spectral resolution observations of the Herbig Ae star HD163296 \thanks{Based on observations collected at the European Southern 
Observatory Paranal, Chile (ESO programme 089.C-0443(A)). }}
\author{R. Garcia Lopez \inst{1,4} \and L.V. Tambovtseva \inst{1,2} \and D. Schertl \inst{1} \and V.P. Grinin \inst{1,2,3} \and K.-H Hofmann \inst{1} \and G. Weigelt \inst{1} \and A. Caratti o Garatti \inst{1,4}  }

\offprints{rgarcia@cp.dias.ie}

\institute{Max-Planck-Institut f\"{u}r Radioastronomie, Auf dem H\"{u}gel 69, D-53121 Bonn, Germany \and Pulkovo Astronomical Observatory of the Russian Academy of Sciences, Pulkovskoe shosse 65, 196140, 
St. Petersburg, Russia \and The V.V. Sobolev Astronomical Institute of the St. Petersburg University, Petrodvorets, 198904 St. Petersburg, Russia \and  Dublin Institute for Advanced Studies, 
31 Fitzwilliam Place, Dublin 2, Ireland }

%
\date{Received date; Accepted date}
%
%
%
\titlerunning{AMBER-HR observations of HD\,163296}
\authorrunning{Garcia Lopez, R. et al.}

\abstract
{Accretion and ejection are tightly connected and represent the fundamental mechanisms regulating star formation. However, the exact physical processes involved are not yet fully understood.
}
{We present high angular and spectral resolution observations of the \brg\ emitting region in the Herbig Ae star HD\,163296 (MWC\,275) in order to probe the origin of this line and constrain the physical 
processes taking place at sub-AU scales in the circumstellar region. }
{By means of VLTI-AMBER observations at high spectral resolution (R$\sim$12\,000), we studied interferometric visibilities, wavelength-differential phases, and closure phases across the 
\brg\ line of HD\,163296. To constrain the physical origin of the \brg\ line in Herbig Ae stars, all the interferometric observables were compared with the predictions of a  
line radiative transfer disc wind  model.}
{The measured visibilities clearly increase within the \brg\ line, indicating that the \brg\ emitting region is more compact than the continuum. By fitting a geometric Gaussian model to the 
continuum-corrected \brg\ visibilities, we derived a compact radius of the \brg\ emitting region of $\sim$0.07$\pm$0.02\,AU (Gaussian half width at half maximum; or a ring-fit radius of $\sim$0.08$\pm$0.02\,AU). 
To interpret the observations, we developed a magneto-centrifugally driven disc wind model. 
Our best disc wind model is able to reproduce, within the errors, all
the interferometric observables and it predicts a launching region with an outer radius of $\sim$0.04\,AU. However, the intensity distribution of the entire disc wind emitting region extends up to 
$\sim$0.16\,AU.}
{Our observations, along with a detailed modelling of the \brg\ emitting region, suggest that most of the \brg\ emission in HD\,163296 originates from a disc wind with a launching region that is over five times 
 more compact than previous estimates of the continuum dust rim radius.}

\keywords{stars: formation -- stars: circumstellar matter -- ISM: jets and outflows -- ISM: individual objects: MWC275, HD163296 -- Infrared: ISM -- techniques: interferometric}

\maketitle

%

\section{Introduction}

As a general view, the accretion and ejection processes in young stellar objects (YSOs) proceed as part of the matter in the disc is lifted up along the magnetospheric accretion columns and accretes 
onto the stellar surface, whereas a small amount of the accreting matter is instead accelerated and collimated outwards in the form of protostellar winds or jets. Accretion and ejection are tightly
 connected and represent the fundamental mechanisms regulating the formation of a star.
The first indications of such a connection 
were found in Classical T Tauri stars (CTTSs) and Herbig AeBe stars through the discovery of a correlation between the luminosity of jet line tracers (such as  the \OI\,6300\,\AA\ line)
 and the infrared excess luminosity \citep{hartigan95,corcoran98}. 
The exact physical processes driving and connecting the accretion and ejection activity are, however, not fully understood.
One of the main reasons is the complex structure of the accretion-ejection region and the small angular scales involved: within less than 1\,AU from the central source (i.e.  $<$10\,mas at a distance of 100\,pc) 
emission from the 
 hot gaseous disc, outflowing material, and the accretion columns is expected. 
In this context, near-infrared interferometry and, in particular, HI Br$\gamma$ spectro-interferometric observations, have become an essential tool with which to investigate the hot and dense gas  
at sub-AU scales.
The first \brg\ spectro-interferometric observations of Herbig AeBe stars showed that in some sources the bulk of the \brg\ emission arises from an extended component \citep[probably 
tracing a disc or stellar wind; e.g.][]{malbet07,eisner_herbig07, kraus08, weigelt11}, in contrast with the scenario in which the \brg\ line would be related with magnetospheric accretion, and so 
unresolved by current 
IR interferometers. 
Despite its unclear origin, the \brg\ line luminosity is directly correlated with the accretion luminosity through empirical relationships \citep{muzerolle98,calvet04} and extensively used to estimate the mass accretion rates of YSOs with different masses 
(0.1--10\,\msun) and at different evolutionary stages  \citep[10$^5$--10$^7$\,yr; e.g.][]{rebeca06,simone11,ale12}.

In order to further constrain the origin of the \brg\ line in Herbig AeBe stars and probe the accretion-ejection connection, we have started a detailed high spectral resolution (HR) VLTI/AMBER study of Herbig AeBe stars. 
We present our results on the Herbig Ae star HD\,163296 (MWC275).
This star is one of the best candidates with which to perform a high angular and spectral study of the accretion-ejection region: HD163296 is located at just $\sim$119\,pc \citep{vanleeuwen07}, it is among the few Herbig Ae stars driving a well collimated  
bipolar jet  \citep[HH409; see  e.g.][]{wassel06}, and it is very likely a single star \citep{swartz05, montesinos09}. The main HD\,163296 stellar parameters  
are listed in 
Table\,\ref{tab:stellar_parameters}.

\begin{table*}[t]
\begin{minipage}[t]{\textwidth}
\caption{\label{tab:stellar_parameters} HD\,163296 stellar parameters}
\centering
\renewcommand{\footnoterule}{} 
\renewcommand*{\thefootnote}{\fnsymbol{footnote}}
\begin{tabular}{ccccccc}
\hline \hline
SpT$^{(1)}$ & T$_{eff}^{(2)}$ & d$^{(3)}$           & R$_*^{(2)}$        & M$_*^{(2)}$        & $\dot{M}_{acc}^{(4)}$        & $\dot{M}_{out}^{(5)}$   \\
          & (K)           & (pc)              & (R$_\odot$)      & (M$_\odot$)      & (10$^{-7}$M$_\odot/yr$)            &   (M$_\odot/yr$)  \\
A1V       & 9250          & 119$\pm$11        & 2.3                & 2.2              & 0.8 -- 4.5\footnotemark[1] \footnotetext{\footnotemark[1]
 Range of \macc\ values reported in the literature (see references in table) derived from the luminosity of the \brg\ line.}                           & 5$\times$10$^{-10}$ --2$\times$10$^{-7}$     \footnotemark[2] \footnotetext{\footnotemark[2]  The first and second values correspond to the atomic (\SII, \OI) and molecular (CO) jet/outflow components.}           \\

\hline
\end{tabular}
\tablebib{(1)~\citet{vandenancker98};
(2)~\citet{mendigutia13};
(3)~\citet{vanleeuwen07};
(4) \citet{rebeca06, donehew11,mendigutia13};
(5) \citet{ellerbroek14,klaassen13}
}
\end{minipage}
\end{table*}

The protoplanetary disc around HD\,163296 has been extensively studied \citep[e.g.][]{isella07,benisty10,tilling12,rosenfeld13}.
Recently, ALMA observations have revealed a vertical temperature gradient across the disc, suggesting dust settling and/or migration \citep[see  e.g.][]{deGregorio13,mathews13}. 
Despite the evidence of dust growth, the disc  still has enough gas content to drive a CO disc wind and a large-scale collimated bipolar jet \citep{wassel06,klaassen13}.  

\begin{figure}
        \centering
        \resizebox{0.78\columnwidth}{!}{\includegraphics{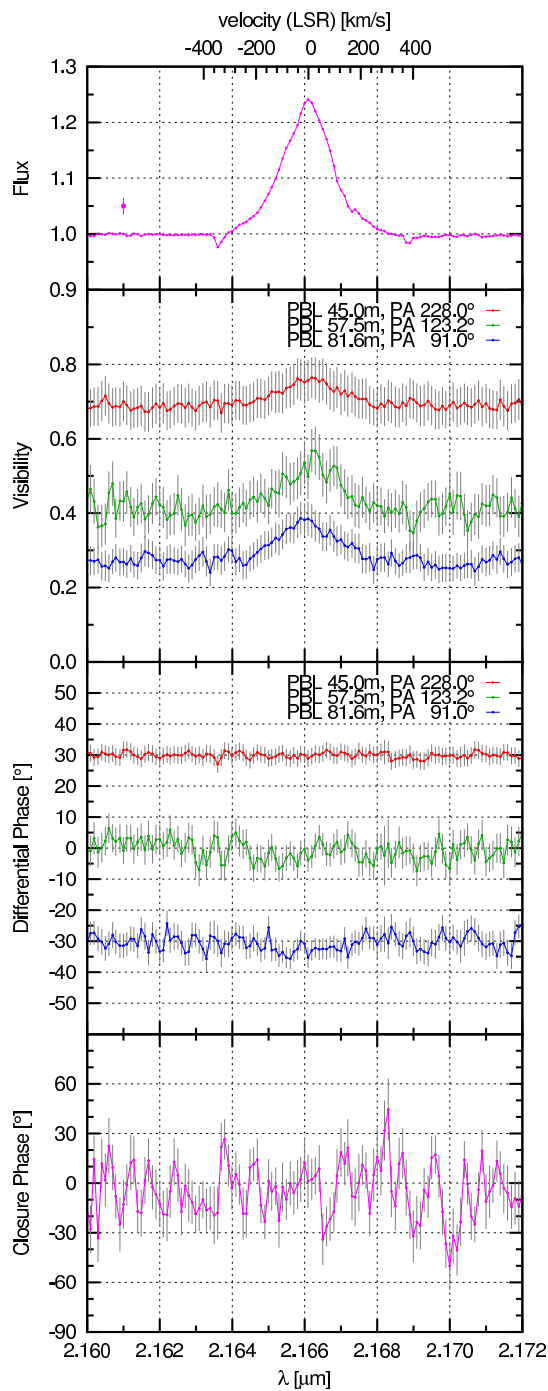}}
                \caption{\label{fig:total_v} AMBER observation of HD~163296 with three different baselines 
                (45.0\,m: red upper line, 57.5\,m: green middle line, and 81.6\,m: blue lower line)  
                and spectral resolution R=12\,000. 
                From  top to bottom: wavelength dependence of flux, visibilities, wavelength-differential phases, and closure phases.
                For clarity, the differential phases of the first and last baselines are shifted by +30 and $-30\degr$, respectively. The  results shown correspond to the averaged value over 
                three different observations (see Sect.\,\ref{sect:obs} and Appendix for details). 
                The wavelength scale is with respect to the local standard of rest (LSR) and corrected by a cloud velocity of $\sim$8\,\kms\ \citep{isella07}.
                }
\end{figure}
Previous, spectro-interferometric studies of HD\,163296 successfully resolved the inner edge of the circumstellar disc, locating it at distance values from the star ranging from 
$\sim$0.19\,AU to $\sim$0.45\,AU 
\citep[e.g.][]{renard10,benisty10,eisner10, eisner14,tannirkulam08a, tannirkulam08b}. It is within this region that the extended \brg\ emission is being emitted \citep{kraus08, eisner14}, although, 
the exact origin of the line emission, and the nature of the innermost gaseous disc are still unclear. 

In the following, we will present our VLTI/AMBER HR observations (R$\sim$12\,000) across the \brg\ line of HD\,163296. The observations, data reduction, and calibration, as well as the interferometric 
observables derived from our observations are presented in Sect.\,\ref{sect:obs} and Sect.\,\ref{sect:observables}. In Sects.\,\ref{sect:line_v}, \ref{sect:disc_model}, and \ref{sect:discussion}, 
a description of the models used to interpret our results and their comparison with the observations are shown. Finally, in Sect.\,\ref{sect:conclusions}, a summary of our results and the main conclusions 
are presented. 

\section{Interferometric VLTI/AMBER/FINITO observations with spectral resolution of 12000}
\label{sect:obs}

We observed HD\,163296 during three nights in May and June 2012 with the ESO Very Large Telescope Interferometer (VLTI)
and its AMBER beam combiner instrument \citep{petrov07}.
For these observations, we used the three 8\,m Unit Telescopes UT2, UT3, and UT4 with AMBER in the high spectral resolution mode (R=12\,000).
The observation parameters are described in Table\,\ref{tab:obs}.
To obtain interferograms with a high signal-to-noise ratio (S/N), we used the fringe tracker FINITO for cophasing and a detector integration time of 6.0\,s per interferogram.

For data reduction, we used our own data reduction software based on the P2VM algorithm \citep{tatulli07} to derive wavelength-dependent visibilities, wavelength-differential phases, and closure phases. 
Along with the science observations of HD~163296, we observed the interferometric calibrator stars HD~163955, HD~162255, and HD~160915, 
which were used to calibrate the transfer function. 
The fringe tracking performance of FINITO is usually different during  the target and calibrator observations.
Therefore, we used all published archival low-resolution AMBER observations of HD\,163296 to calibrate the continuum visibilities. 
The errors of the K-band squared visibilities are approximately $\pm$5\% or slightly better during the best nights.
Each of our three recorded datasets consists of 75 target and calibrator interferograms. 
Because all three observations were performed with very similar projected baseline lengths and position angles (PAs), we averaged the results of all three observations in order to get a better S/N per 
spectral channel 
(Fig.\,\ref{fig:total_v}). 
The smaller errors of the averages of the three datasets were obtained from the errors of the individual results in the standard way.
Because the absolute continuum visibilities were calibrated using the published archive low resolution (LR) observations, in the next step, we took the errors of these LR archival data into account and we calculated the
total error in the standard way.
The three individual results are shown in the Appendix (Fig.\,\ref{fig:appendix_total_v}).
The wavelength calibration was accomplished using the numerous telluric lines present in the region 2.15--2.19\,\um\ 
\citep[see][for more details on the wavelength calibration method]{weigelt11}. We estimate an uncertainty in the wavelength calibration of $\sim 3~$\,\kms.

\section{Interferometric observables}
\label{sect:observables}

Our VLTI/AMBER observations provide four direct observables: \brg\ line profile, visibilities, differential phases, and closure phases. 
These observables allow us to retrieve information about the size and kinematics of the \brg\ emitting region. 

Figure\,\ref{fig:total_v} shows  (from top to bottom) the \brg\ line profile, wavelength visibilities, differential phases, and closure phase of our \brg\ spectro-interferometric observations of HD\,163296. 
The wavelength-dependent visibilities (second panel from top) clearly increase within the \brg\ line in all our baselines. This indicates that the \brg\ emitting region is more compact than the continuum 
emitting region. 

Previous spectro-interferometric studies of HD\,163296 at medium resolution had also detected an increase of the visibility within the \brg\ line \citep{kraus08, eisner14}. 
Our HR mode AMBER observations have, however, allowed us to measure the visibilities and phases in $\sim$30 different spectral channels across the Doppler-broadened Br$\gamma$ line, helping us to better 
distinguish the contribution of the line to that of the continuum.   
The measured differential and closure phases do not clearly show any significant deviation from zero within the error bars of $\pm$5\degr, and $\pm$15\degr, respectively
(Fig.\,\ref{fig:total_v}, bottom panel).

\begin{table*}[]
\begin{minipage}[t]{\textwidth}
\caption{\label{tab:obs} Log of the VLTI/AMBER/FINITO observations of HD~163296 and its calibrators.}
\centering
\renewcommand{\footnoterule}{} 
\begin{tabular}{cccccccccc}
\hline\hline
HD\,163296\footnotetext{Note: Object and calibrators were observed with the same spectral mode, DIT, and wavelength range.}    & \multicolumn{2}{c}{Time [UT]}& Unit Telescope & Spectral & Wavelength  & DIT\footnote{Detector integration time per interferogram.} & N\footnote{Number of interferograms.}  & Seeing  & Calibrator \\
Observation  & Start & End                  & array          & mode\footnote{High spectral resolution mode in the $K$ band using the fringe tracker FINITO.} & range       &         &        &         & index      \\
Date         &       &                      &                &          & [$\mu$m]    & [s]     &        & [arcsec]& (see below)\\
\noalign{\smallskip}
\hline
\noalign{\smallskip}
2012 May  11 & 08:41 & 08:49                & UT2--UT3--UT4    &  HR--K--F  & 2.147--2.194 & 6       & 75     & 0.6--0.8 & (a)+(b) \\
2012 May  12 & 09:20 & 09:28                & UT2--UT3--UT4    &  HR--K--F  & 2.147--2.194 & 6       & 75     & 0.9--1.5 & (c) \\
2012 Jun. 04 & 07:23 & 07:30                & UT2--UT3--UT4    &  HR--K--F  & 2.147--2.194 & 6       & 75     & 0.7--0.9 & (d)+(e) \\
\noalign{\smallskip}
\hline
\end{tabular}

\vspace*{3mm}
\begin{tabular}{cccccccc}
\hline\hline
Calibrator    & Date         & \multicolumn{2}{c}{Time [UT]} & UT array    &  N$^b$ & Seeing  & Uniform-disc \\
Name          &              & Start & End                   &             &        &         & diameter\footnote{ UD diameter taken from JMMC Stellar Diameters Catalogue - JSDC \citep{lafrasse10}.} \\
    (index)   &              &       &                       &             &        & [arcsec]& [mas] \\
\noalign{\smallskip} \hline \noalign{\smallskip}
HD\,162255 (a) & 2012 May 11  & 08:09 & 08:17                 & UT2--UT3--UT4 &   75    & 0.6--0.7 & 0.34$\pm$0.02         \\
HD\,160915 (b) & 2012 May 11  & 09:05 & 09:13                 & UT2--UT3--UT4 &   75    & 0.6--0.7 & 0.68$\pm$0.05 \\
HD\,160915 (c) & 2012 May 12  & 08:56 & 09:04                 & UT2--UT3--UT4 &  75    & 1.0--1.5 & 0.68$\pm$0.05 \\
HD\,162255 (d) & 2012 Jun. 04 & 06:35 & 06:42                 & UT2--UT3--UT4 &   75    & 0.9--1.2 & 0.34$\pm$0.02 \\
HD\,163955 (e) & 2012 Jun. 04 & 07:45 & 07:53                 & UT2--UT3--UT4 &   75    & 0.6--0.8 & 0.40$\pm$0.03 \\
\noalign{\smallskip} \hline
\end{tabular}
\end{minipage}

\end{table*}

\section{Geometric modelling: The characteristic size  of the Br$\gamma$ line-emitting region}
\label{sect:line_v}

The high spectral resolution of the AMBER data  allows us to measure the visibilities of the pure Br$\gamma$ line-emitting region for several
spectral channels across the Br$\gamma$ emission line (see Fig.\,\ref{fig:line_v}). 
These continuum-compensated line visibilities (required for the size determination of the Br$\gamma$ line-emitting region) were calculated in the
following way. Within the  wavelength region of the Br$\gamma$ line emission, the measured visibility has two constituents: the pure
line-emitting component and the continuum-emitting component, which includes continuum emission from both the circumstellar environment and the unresolved
central star. The emission line visibility $V_{{\rm Br}\gamma}$ in each spectral channel can be written as
%
%
\begin{eqnarray}\label{eq_brgvis1}
        & & F_{{\rm Br}\gamma}V_{{\rm Br}\gamma} = \nonumber \\
        & & = \sqrt{ |F_{\rm tot}V_{\rm tot}|^2 + |F_{\rm c}V_{\rm c}|^2 - 2\,F_{\rm tot}V_{\rm tot}\,F_{\rm c}V_{\rm c}\cdot cos{\Phi} }\\
        & & = |F_{\rm tot}V_{\rm tot} - F_{\rm c}V_{\rm c}|,
\end{eqnarray}
%
if the differential phase $\phi$ is zero \citep{weigelt07}. The value F$_{Br\gamma}$ is the wavelength-dependent line flux, 
$V_{\rm tot}$ ($F_{\rm tot}$) denotes the measured total visibility (flux) in the  ${\rm Br}\gamma$ line, $V_{\rm c}$ ($F_{\rm c}$) is the 
visibility (flux) in the continuum, and $\Phi$ is the measured wavelength-differential phase within the ${\rm Br}\gamma$ line. 
The intrinsic \brg\ photospheric absorption feature was taken into account in this analysis by considering a synthetic spectrum with the same spectral type and surface gravity 
of HD\,163296 \citep{eisner10, mendigutia13}.

We have derived the pure line visibility only in spectral regions where the line flux is higher than $\sim$10\% of the continuum flux.
The results are shown in Fig.\,\ref{fig:line_v}. 
The average line visibility is $\sim$1 at the shortest baselines (45\,m), $\sim$0.9 at medium baselines (57.5\,m), and $\sim$0.8 at the longest baselines (81.6\,m).
The approximate size of the line-emitting region was obtained by fitting a circular symmetric Gaussian model to the line visibilities presented in Fig.\,\ref{fig:line_v}. We obtained a Gaussian half
width at half maximum (HWHM) radius of 0.6$\pm$0.2\,mas or $\sim$0.07$\pm$0.02\,AU. 
Similarly, by fitting a ring model with a ring width of 20\% of the inner ring radius, an inner radius of 0.7$\pm$0.2\,mas or $\sim$0.08$\pm$0.02\,AU, is found. This radius is smaller than the inner
continuum dust rim radius of R$_{rim}\sim$0.19--0.45\,AU reported in the literature \citep{tannirkulam08a,tannirkulam08b, benisty10, eisner14}.

\section{Disc and wind model}
\label{sect:disc_model}

Following our previous work on the Herbig Be star MWC\,297, we developed a disc wind model to explain the observed \brg\ line profile, line visibilities, differential phases, and 
closure phase \citep{weigelt11,grinin11,larisa14}. 
A detailed description of the disc wind model, the complete algorithm for the model computation, and a detailed description on how the disc wind model solution 
depends on the kinematic and
physical parameters can be found in these papers. In the following, we will briefly summarise the 
main characteristics of our disc wind model.

The disc wind model employed here is based on the magneto-centrifugal disc wind model of \cite{blandford82}, adapted for the particular case of Herbig AeBe stars. 
For simplicity, the disc wind consists only of hydrogen atoms with constant temperature ($\sim$10\,000\,K) along the wind streamlines. This approximation is in agreement with the so-called warm 
disc wind models \citep{safier93,garcia01a}, in which the wind is rapidly heated by ambipolar diffusion to a temperature of $\sim$10\,000\,K. 
In these models, the wind electron temperature in the acceleration zone near the
disc surface  is not high enough to excite the \brg\ line emission. Therefore, in our model, the low-temperature region below a certain height value does not contribute to the \brg\ disc wind emission.

For the calculations of the model images of the emitting region in the line frequencies and their corresponding interferometric quantities (i.e.  visibilities and phases), 
we used the same approach as in \cite{weigelt11}. For the 
calculations of the ionization state and the number densities of the atomic levels, we adopted the numerical codes developed by \cite{grinin90} and \cite{larisa01} for moving media. These codes are based 
on the Sobolev approximation \citep{sobolev60} in combination with
the exact integration of the line intensities (see  Appendix A in \citealt{weigelt11}).
This method takes into account the radiative coupling in the local environment of each point caused by multiple scattering. The calculations were made for the 15-level model of the hydrogen atom plus continuum, 
taking into account both collision and radiative processes of excitation and ionization (see  \citealt{grinin90} for more details). 

Very briefly, our disc wind model considers a wind launched from an inner radius $\omega_1$ to an outer radius $\omega_N$ (wind footpoints). 
The disc wind half opening angle ($\theta$) is defined as the angle between the innermost wind streamline and the system axis. 
The angle range (30\degr--45\degr) in Table\,\ref{tab:model_parameters} is consistent with the \cite{blandford82} disc wind solution.   
The local mass-loss rate per unit area on the disc 
surface is defined as $\dot{m}(\omega)\sim \omega^{-\gamma}$, where $\gamma$ is the so-called mass-loading parameter that controls the ejection efficiency. Thus, the total mass-loss rate is given by
\begin{equation}
\dot M_{w} = 2\int\limits_{\omega_1}^{\omega_N}\dot
m_{w}(\omega)\,2\,\pi\,\omega\,d\omega.
\end{equation}

In all our disc wind models, the accretion disc is assumed to be transparent to radiation at radii up to the dust sublimation radius. At larger radii, the disc is an opaque screen that shields a large 
fraction of the wind emission because the observer cannot see the disc wind emission on the back side of the disc. It should be noted that this assumption (i.e. opaque disc with an inner gap) is only 
valid if the mass accretion rate (\macc) does not exceed a value of $\sim$10$^{-6}$\,\msyr\ \citep{muzerolle04}.

All free model parameters are listed in Table\,\ref{tab:model_parameters}. Some model parameters such as $\dot{M}_w$, the stellar parameters and the disc inclination angle were taken from the literature 
(see Tables\,\ref{tab:model_parameters}  and 
\ref{tab:stellar_parameters}). The modelled total mass-loss rate ($\dot{M}_w$\footnote{From now on, we use $\dot{M}_w$ for the modelled total mass-loss rate and $\dot{M}_{out}$ for the observed
value from the literature.}) was allowed to vary within $\sim$0.1-1.0 times the \macc\ observed average value (see Table\,\ref{tab:stellar_parameters}). 
In addition, the estimated value of the corotation radius is set as a lower limit to the inner disc wind launching radius value $\omega_1$ (see Sect.\,\ref{sect:discussion}).
These parameters, along with our interferometric observations further constrain our disc wind model solution. 
For instance, the measured visibilities constrain the size of the disc wind launching region, 
and the adopted $\dot{M}_w$ and measured line profile constrain the wind dynamics \citep[see][for a complete discussion]{larisa14}.

Finally, in order to compute the model differential phases, a continuum model consisting of two components, an inner and an outer disc, was adopted \citep[e.g.][]{tannirkulam08a, benisty10}.
This model is able to reproduce both the observed continuum visibilities and continuum flux of HD\,163296. 
The dominant outer disc is a temperature-gradient disc with a power law index of -0.50, inner temperature of 1500\,K, and a dust rim radius of 20\,R$_*$ (i.e. $\sim$0.21\,AU). 
The stellar flux is 0.14 of the observed flux in the K-band \citep{benisty10}. 
The additional optically thin inner disc has an inner radius of 5\,R$_*$, an outer radius equal to the inner radius of the outer disc (i.e.  20\,R$_*$), 
and a constant intensity distribution flux of 20\% of the observed continuum flux.

\begin{table*}
\begin{minipage}[t]{\textwidth}
\caption{\label{tab:model_parameters} Disc wind model parameters} 
\centering
\renewcommand{\footnoterule}{} 
\begin{tabular}{ccccc}
\hline \hline
  Parameters\footnote{The definition of the listed disc wind model parameters is described in detail in \cite{weigelt11}.} 
                &        Range\footnote{Disc inclination and position angle (major axis) of $\sim$38\degr\ and $\sim$138\degr, respectively, were assumed \citep[gaseous ALMA disc;][]{deGregorio13}, 
as well as the stellar parameters listed in Table\,\ref{tab:stellar_parameters}.}                        &  MW6$^b$       & MW26 & MW47         \\
\hline
 Temperature\,(K)                  & 8000 -- 10\,000                                                         & 10\,000        & 10\,000   & 10\,000         \\
 Half opening angle ($\theta$)                  & 30$^\circ$ -- 45$^\circ$                               & 45$^\circ$     & 30$^\circ$   &  20$^\circ$     \\
 Inner radius\,($\omega_1$(R$_*$))             & 2 -- 3                                                  & 2.0 (0.02\,AU) & 2.0 (0.02\,AU)  & 5.0 (0.05\,AU)  \\
 Outer radius\,($\omega_N$(R$_*$))            & 4 -- 30                                                  & 4.0 (0.04\,AU) & 15.0 (0.16\,AU) & 15.0 (0.16\,AU)  \\
 Acceleration parameter\,($\beta$)                    & 1 -- 7                                           & 5              & 3               & 4                 \\
 Mass load parameter\,($\gamma$)                    & 1 -- 5                                             & 3                  & 3               & 3         \\           
 Mass loss rate\,($\dot{M}_w$(M$_\odot$/yr)) & $10^{-8}-10^{-7}$                                         & 5$\times$10$^{-8}$  &  5$\times$10$^{-8}$ &  5$\times$10$^{-8}$      \\
\hline
\end{tabular}
\end{minipage}
\end{table*}

\begin{figure}[!t]
        \centering
        \resizebox{0.78\columnwidth}{!}{\includegraphics{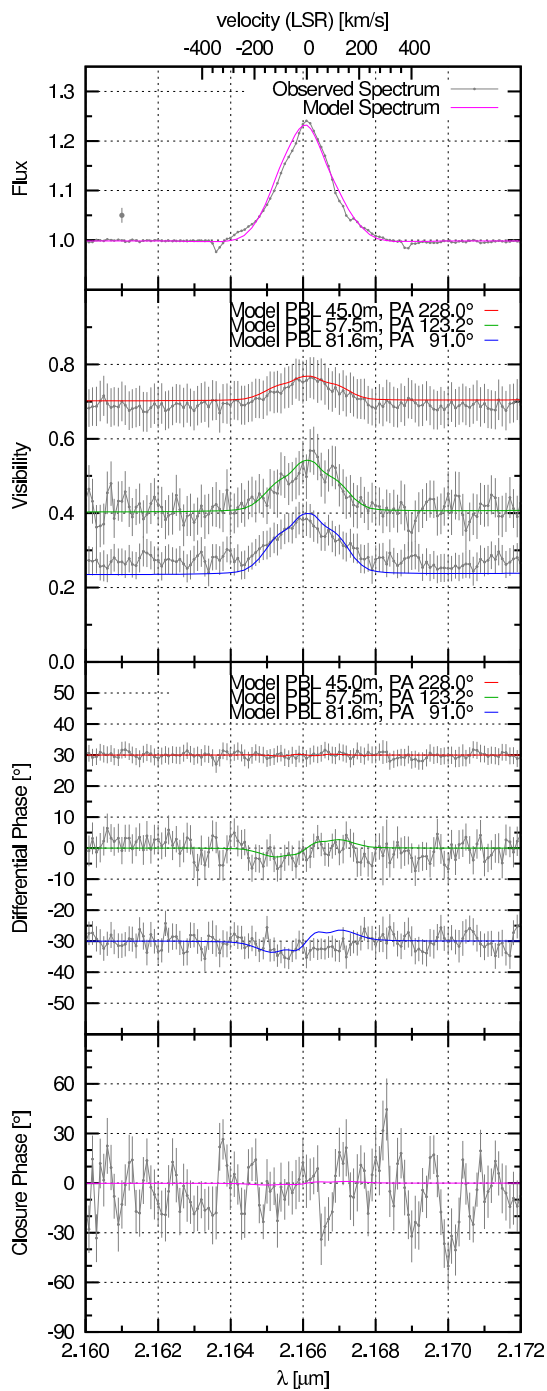}}
        \caption{\label{fig:best_fit} Comparison of our interferometric observations with the interferometric quantities derived from our best disc wind model MW6.
                                      From top to bottom, observed \brg\ line profile (grey) and model line profile (pink), observed visibilities (grey dots with error bars) and model visibilities 
                                      (coloured lines), observed and model wavelength-differential phases, and, observed and model closured phases.}       
\end{figure}

\section{Results}
\label{sect:discussion}

In order to find a disc wind model that is able to approximately reproduce all interferometric observables, 
we calculated more than 100 models within the parameter ranges listed in Table\,\ref{tab:model_parameters}.
All our models were computed assuming a disc inclination (angle between the line of sight and the system axis of the gaseous ALMA gas disc) of $\sim$38\degr\ and a 
position angle of the major axis of the disc of $\sim$138\degr\  \citep{deGregorio13}. 
We computed the intensity distributions of the models for all the wavelength channels across the \brg\ line. 

Because our observations do not provide us with wavelength-dependent images (as the modelling does), but only with the \brg\ line profile
and information on visibilities and phases, we first had to derive the following interferometric quantities from the wavelength-dependent model intensity distributions:
 model line profiles, model visibilities, and differential and closure phases of the model.
All the interferometric model quantities were computed for the same baseline lengths, baseline PAs, and spectral resolution as the observations.
This method allowed us to compare the observed interferometric quantities (i.e.  line profile, visibilities, differential phases, and closure phases) with the corresponding interferometric model quantities 
and to find out which of our models can best reproduce the observations.

Figures\,\ref{fig:best_fit} and \ref{fig:line_v} show a comparison of the observations with the interferometric quantities derived from our best disc wind model MW6 (see Table\,\ref{tab:model_parameters}
for a list of the model parameters). 
The intensity distribution maps of this model at several radial velocities are presented in Fig.\,\ref{fig:intensity_distribution}.
All model parameters are listed in Table\,\ref{tab:model_parameters}. 
As shown in the figures, model MW6 approximately agrees with the observed line profile (Fig.\,\ref{fig:best_fit}, upper panel), visibilities (Fig.\,\ref{fig:best_fit}, middle panel), 
and small differential phases and closure phases 
(Fig.\,\ref{fig:best_fit}; the measured differential and closure phases are approximately zero within our errors; the model closure phase is $\sim$1\degr).
Moreover, this model is also able to approximately reproduce the visibilities and small differential phase of previous Keck interferometric observations of 
this source \citep{eisner10,eisner14}.
However, it slightly overpredicts the pure \brg\ line visibilities for the longest baseline (see Fig.\,\ref{fig:line_v}).
This can also be seen in Fig.\,\ref{fig:best_fit}, where the model continuum visibility is slightly below the observed value for the longest baseline.

In order to obtain a lower value of the modelled pure line visibility at the longest baseline, it would be necessary to increase the size of the disc wind region (e.g. $\omega_N$).
Two examples of such models (i.e. with $\omega_N>$4\,R$_*$) can be found in Fig.\,\ref{fig:appendix_largeWN1} (model MW26) and Fig.\,\ref{fig:appendix_largeWN2} (model MW47). 
Model MW26 has the same 
$\omega_1$ value as our best model MW6, but a $\omega_N$ value almost 4 times larger (0.16\,AU). Model MW47 has larger values of $\omega_1$ (0.05AU) and $\omega_N$ (0.16\,AU). 
The modelled interferometric results 
presented in Figs.\,\ref{fig:appendix_largeWN1} and Figs.\,\ref{fig:appendix_largeWN2} show that these larger disc wind launching radii have little influence on the line visibility. Both new 
models have, however, large differential phases ($\gtrsim$34\degr) that disagree with the observed differential phases (approximately zero within the average error of $\sim\pm$5\degr). These large differential 
phases
are caused by the
larger radii, that produce larger photo centre shifts with respect to the continuum. 
It should be noted that the acceleration parameters of models MW6, MW27, and MW47, as well as the half opening angles, differ slightly  from model to model ($\beta$=5, 3, and 4, and $\theta$=45\degr, 
30\degr, and 20\degr, respectively).
However, these model parameters mainly modify the \brg\ line profile \citep[see e.g.][]{larisa14}, and have little effect on the line visibilities and differential phases which mostly depend on the 
$\omega_1$ and $\omega_N$ parameters.
Therefore, a good match
between our interferometric observables (line profile, visibilities, and differential and closure phases) and the modelled results can only be  achieved when combining lower values of the inner 
and outer disc wind footpoints, that is, when a compact disc wind is considered.

\section{Discussion}

\begin{figure}
        \centering
        \resizebox{0.78\columnwidth}{!}{\includegraphics{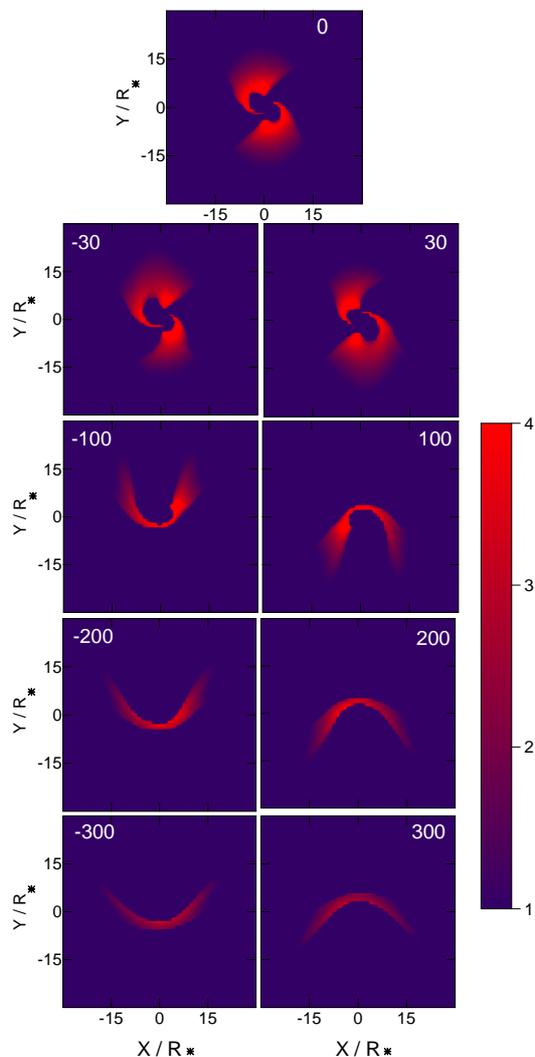}}
        \caption{\label{fig:intensity_distribution}
                \brg\ intensity distribution maps of our best disc wind model MW6 (see Table\,\ref{tab:model_parameters}) at several radial velocities, as indicated by the white labels  
                in units of \kms.
                Emission from the disc continuum and the central star is not shown.  The colours represent the intensity in logarithmic scale in arbitrary units.
        }        
\end{figure}

The disc wind launching region of our best model MW6 is quite compact, extending from an inner radius of $\sim$0.02\,AU to an outer radius of $\sim$0.04\,AU. 
This latter value is not far from the lowest outer launching radius of 0.03\,AU derived from \cite{ferreira06} from a comparison of optical jet observations of TTauri stars 
(typical R$_{cor}\sim$0.1\,AU) and the predictions from extended disc wind models. Even if a direct comparison between TTauri and Herbig AeBe stars is difficult, this result suggests that disc wind models with 
low $\omega_N$ values can explain some observations of extended emission in YSOs.
However,  Fig.\,\ref{fig:intensity_distribution} shows that the entire disc wind emitting region extends to an outer radius of $\sim$15\,R$_*$ or $\sim$0.16\,AU. 
As discussed in the previous section, this allows us to obtain approximately the required low visibilities at the longest baseline, as well as small differential phases, and to fit the \brg\ line profile.

It should be noted that our modelling does not exclude that a fraction of the \brg\ flux might be emitted from a compact magnetosphere, a stellar wind, and/or an X-wind 
\citep[see  e.g.][for examples of hybrid models]{larisa14}. 
Because of the small magnetic fields measured in Herbig AeBe stars,
the distance at which the circumstellar disc of Herbig AeBe stars is truncated by the stellar magnetic field (the truncation radius) is expected to be much smaller
than in CTTSs and it is located within the corotation radius (R$_{cor}$) \citep{shu94,muzerolle04}. 
For the case of HD\,163296, the corotation radius, R$_{cor}$, is $\sim$1.6\,R$_*$ or $\sim$0.02\,AU\footnote{
The corotation value of 1.6\,R$_*$ was derived from the stellar parameters presented in Table\,\ref{tab:stellar_parameters}, assuming an inclination angle of 38$^\circ$ 
\citep[][see Table\,\ref{tab:model_parameters}]{deGregorio13}, 
and a projected rotational velocity of $v \sin i$  = 133 km s$^{-1}$ \citep{montesinos09}.}. 
Therefore, the magnetosphere outer radius ($\lesssim$0.02\,AU) is much smaller than the outer radius of the whole disc wind emitting region ($\sim$0.16\,AU; see Fig.\,\ref{fig:intensity_distribution}), 
and comparable to the inner launching radius parameter $\omega_1$ in our disc wind model MW6. As a consequence, 
any additional emission in our model located within the corotation radius ($\sim$0.02\,AU) would increase the modelled \brg\ visibilities of the disc wind intensity
distribution. To illustrate this effect, we employed a hybrid model 
(disc wind plus magnetosphere) as described in \cite{larisa14}, and add the contribution of a compact magnetosphere accounting for 40\% of the total \brg\ flux to our best model MW6 (a brief description 
of the magnestosphere parameters and their values can be found in Appendix\,\ref{sect:appendix_hybrid_model}). 
To account for the extra emission from the magnetosphere and obtain a good fit to the line profile, the \brg\ emitting region in model MW6 was reduced by increasing the size of the low-temperature region at 
the wind base where the gas does not produce \brg\ emission (see Sect.\,\ref{sect:disc_model}). In this way, the disc wind \brg\ emission was reduced to a 60\% of the total \brg\ flux.
Figure\,\ref{fig:appendix_MW6+magnetosphere} shows the result of this model overplotted on our observations: the pure \brg\ line visibilities are now $\sim$1 at all  
baselines. To conclude, our hybrid model (i.e. disc wind plus magnetosphere) suggests that the observed low visibility values can only be achieved with larger disc wind footpoint radii than in model MW6, 
but this would in turn produce excessive values of the differential phase as well, contradicting our observations.

Nevertheless, we would like to point out that the aim of this paper is to investigate which mechanism is responsible for most of  
the \brg\ emission in HD\,163296. Therefore, our observations do not rule out that some of the \brg\ emission is produced in an X-wind and/or a magnetosphere in HD\,163296, 
but they  would  probably have to act in combination with a more extended emission like the disc wind 
reported here in order to reproduce the observed \brg\ line visibilities and the line profile.

The presence of a disc wind in HD\,163296 is also supported by ALMA observations which revealed an extended CO rotating disc wind emerging from this source \citep{klaassen13}. 
Our observations also indicate  a compact disc wind component traced by the \brg\ line, and launched well within the dust sublimation radius. 
This, in turn, would explain why no dust has been detected along the long-scale optical/infrared bipolar jet \citep{ellerbroek14}, that is, 
the jet is launched from an inner disc region where none or only little dust is present.

\section{Summary and conclusions}

\label{sect:conclusions}
In this paper, we present AMBER high spectral resolution (R=12\,000) observations of the young Herbig Ae star HD\,163296. The high spectral resolution has allowed us to sample the interferometric observables
(i.e.  line profile, visibilities, differential phases, and closure phases)
in many  spectral channels across the \brg\ line.
Our observations  show that the visibility within the \brg\ line is higher than in the continuum 
(Fig.\,\ref{fig:total_v}). Therefore, the \brg\ emitting region is less extended than the continuum emitting region. To characterise the size of the \brg\ emitting region, 
we fitted geometric Gaussian and ring models to the derived continuum-corrected \brg\ line visibilities (i.e.  the pure \brg\ visibilities). We derived a HWHM Gaussian 
radius of 0.6$\pm$0.2\,mas or $\sim$0.07$\pm$0.02\,AU, and ring-fit radius of 0.7$\pm$0.2\,mas or $\sim$0.08$\pm$0.02\,AU for the \brg\ line-emitting region (for the adopted distance of 119\,pc). 

To obtain a more physical interpretation of our AMBER observations, we employed our own line radiative transfer disc wind model 
\citep{weigelt11, grinin11}.  
Using this approach, we computed a model that approximately agrees with all interferometric observables. 
Figures\,\ref{fig:line_v} and \ref{fig:best_fit} show that our best disc wind model MW6 (Table\,\ref{tab:model_parameters}) agrees with 
(1) the \brg\ line profile,
(2) the total visibilities across the \brg\ line, 
(3) the continuum-corrected \brg\ line visibilities, and 
(4) the small observed wavelength-differential phases and closure phases.
Therefore, our AMBER HR spectro-interferometric observations of the \brg\ line, along with a detailed modelling of the line-emitting region suggest that the \brg\ line-emitting region in HD\,163296 mainly
 originates from a disc wind with an inner launching radius of 0.02\,AU and an outer launching radius of 0.04\,AU. However, Fig.\,\ref{fig:intensity_distribution} shows that the entire disc wind 
 emitting region extends to an outer radius of $\sim$15\,R$_*$ or $\sim$0.16\,AU. The modelled disc wind emitting region is more compact than the inner continuum dust rim radius
reported in the literature \citep[R$_{rim}\sim$0.19-0.45\,AU;][]{benisty10, eisner10, eisner14,tannirkulam08a,tannirkulam08b}. Our modelling does not exclude that some of the \brg\ emission is produced in the magnetosphere. 
 However, in addition to the emission from the magnetosphere, emission from a more extended region, like the disc wind presented here, would be required to explain the interferometric observations.
 
\begin{acknowledgements}
The authors thank the Paranal science operation team for carrying out these observations in service mode. R.G.L and A.C.G. were supported by the Science Foundation of
Ireland, grant 13/ERC/I2907. L.V.T. and V.P.G were supported by grant of the Presidium of RAS P41.
We thank the referee for his/her useful comments and suggestions, which helped to improve the paper.
\end{acknowledgements}

\bibliographystyle{aa}
\bibliography{references}

\Online

\begin{appendix}

\section{Individual set of observations}

\begin{figure*}
\centering
        \includegraphics[width= 13.5cm, angle =0]{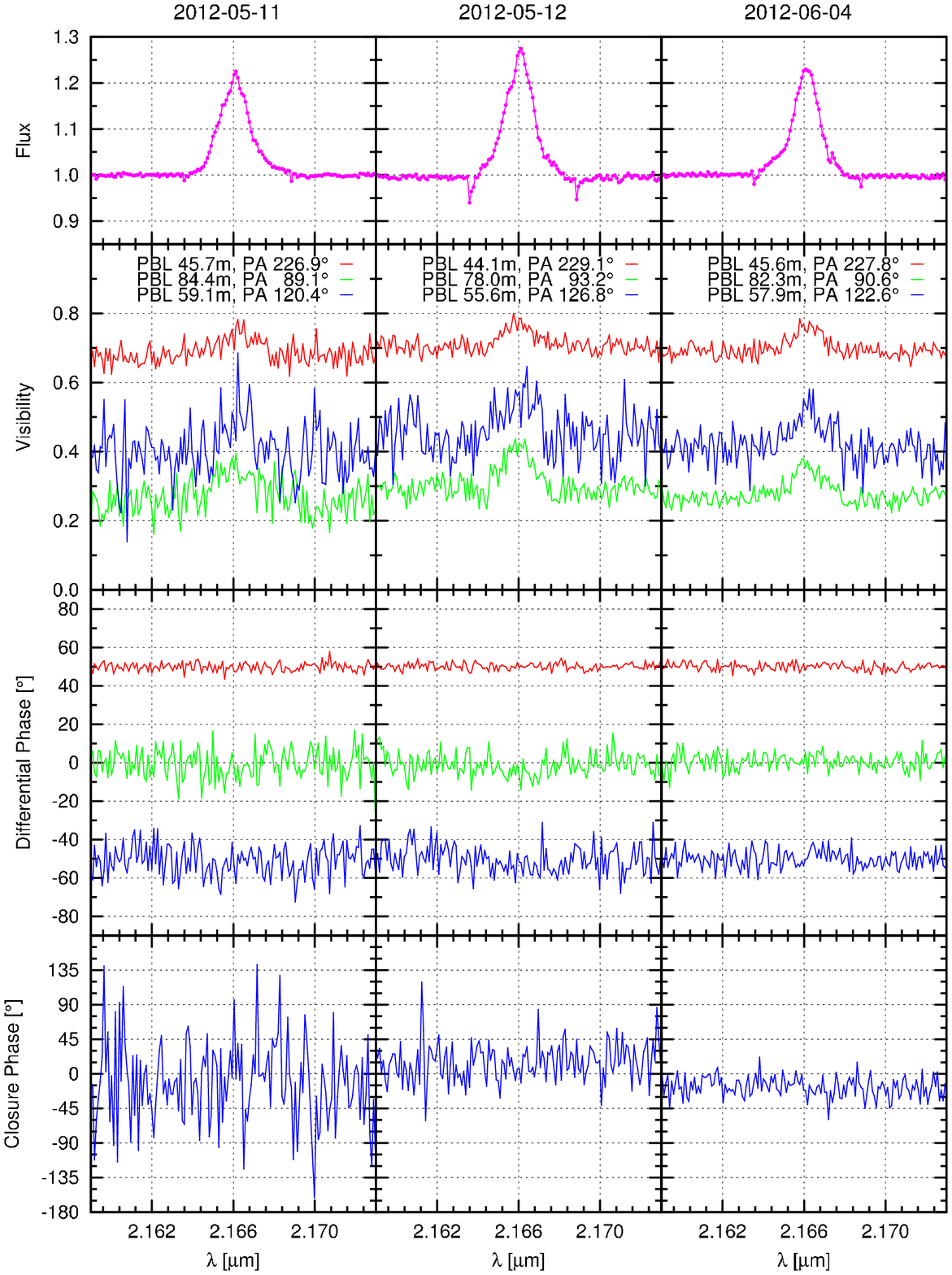}
        \caption{\label{fig:appendix_total_v} All three AMBER observations of HD~163296 with spectral resolution of 12\,000. 
                The three columns show the observed intereferometric results from 2012 May 11 (left), 2012 May 12 (middle), and 2012 June 04 (right).
                Shown from  top to bottom are wavelength dependence of flux, visibilities, wavelength-differential phases (for better visibility, the differential phases of the first and last
                baselines are shifted by $+50\degr$ and $-50\degr$, respectively), and closure phase observed at projected baselines as shown in the plot.
                The wavelength scale at the bottom is corrected to the local standard of rest. The typical visibilities, differential, and closure phases errors are $\pm$5\%, 5\degr, and 15\degr.
                }
\end{figure*}

\section{Pure line visibilities}

\begin{figure}
        \centering
        \resizebox{0.78\columnwidth}{!}{\includegraphics{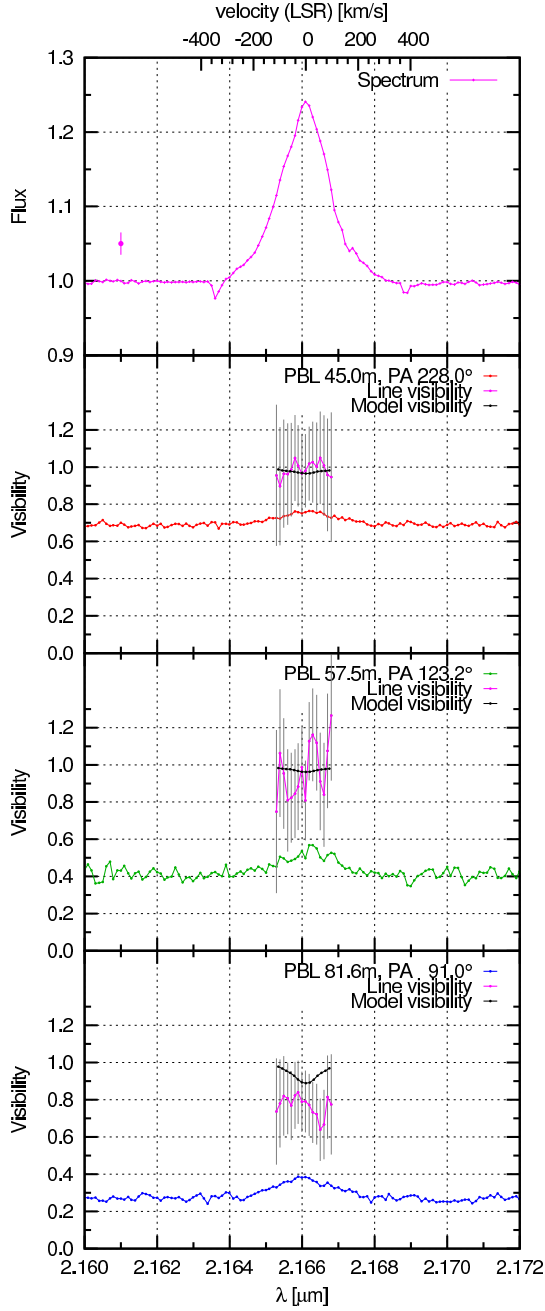}}
        \caption{\label{fig:line_v}
                Comparison of the observed and modelled pure Br$\gamma$ line visibilities of our AMBER observation of HD~163296.
                From  top to bottom: wavelength dependence of flux, visibilities of the first, second, and third baseline.
                In each visibility panel: (1) the observed total visibilities (red, green, blue, as in Fig.\,\ref{fig:total_v}), (2) the observed
                continuum-compensated pure Br$\gamma$ line visibilities (pink), and the modelled pure \brg\ line visibilities (black; model MW6, Table\,\ref{tab:model_parameters}) are shown.
        }       
\end{figure}

\section{Examples of computed disc wind models}

\begin{figure*}
\centering
        \includegraphics[width=0.9\columnwidth]{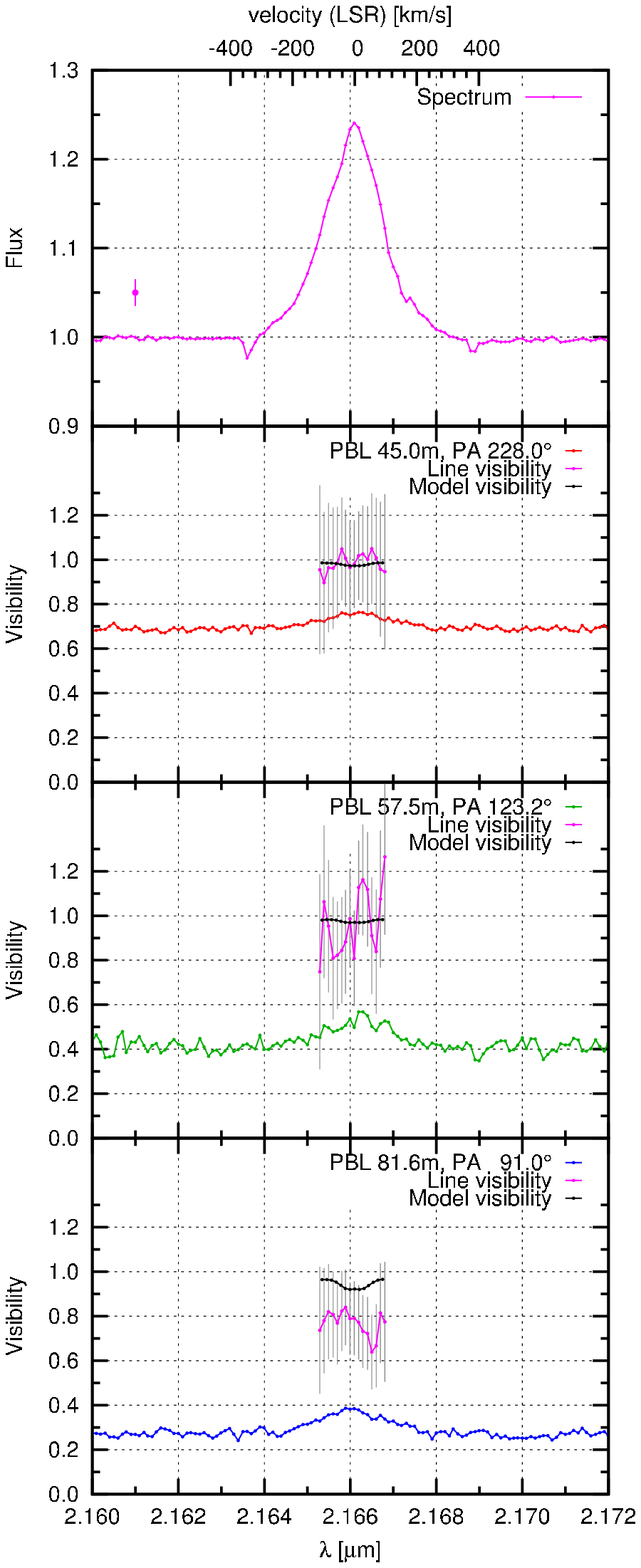}
        \includegraphics[width=0.85\columnwidth]{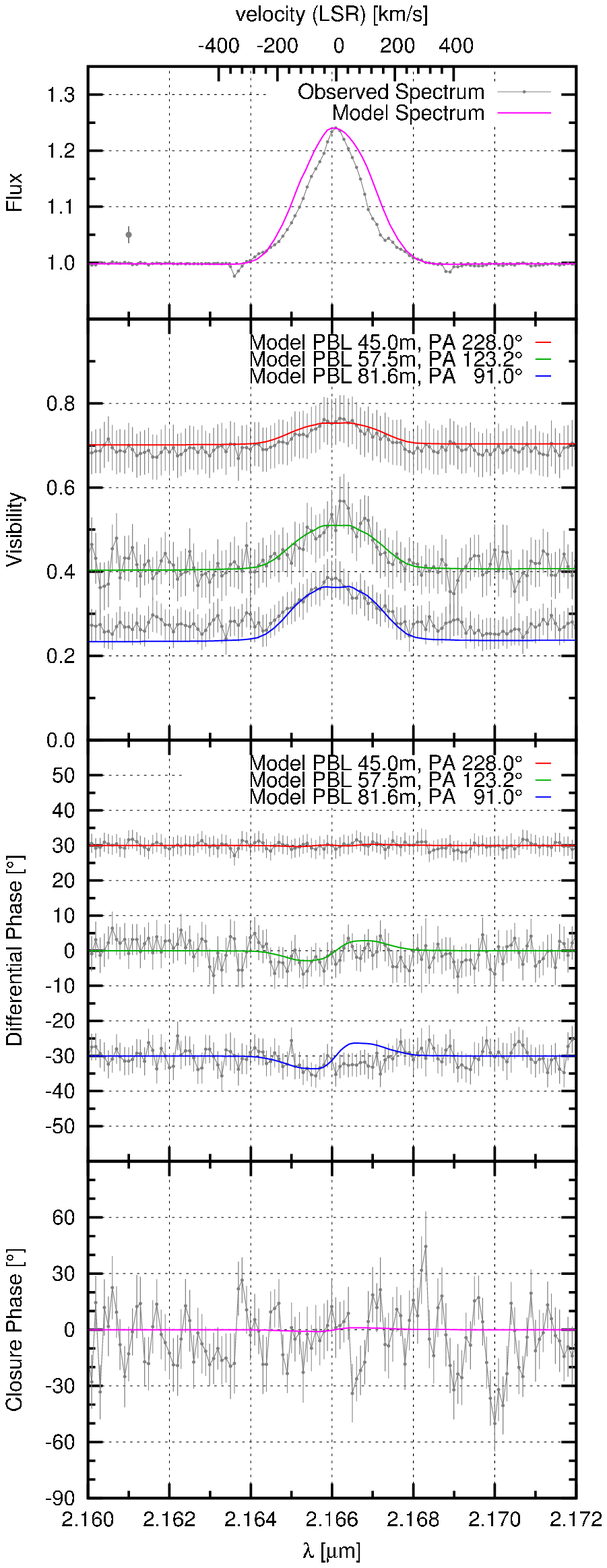}
        \caption{\label{fig:appendix_largeWN1}
                Same as Fig. \ref{fig:line_v} (left panel) and Fig. \ref{fig:best_fit} (right panel) but for model MW26 (see, Table\,\ref{tab:model_parameters}).
        }       
\end{figure*}

\begin{figure*}
\centering
        \includegraphics[width=0.9\columnwidth]{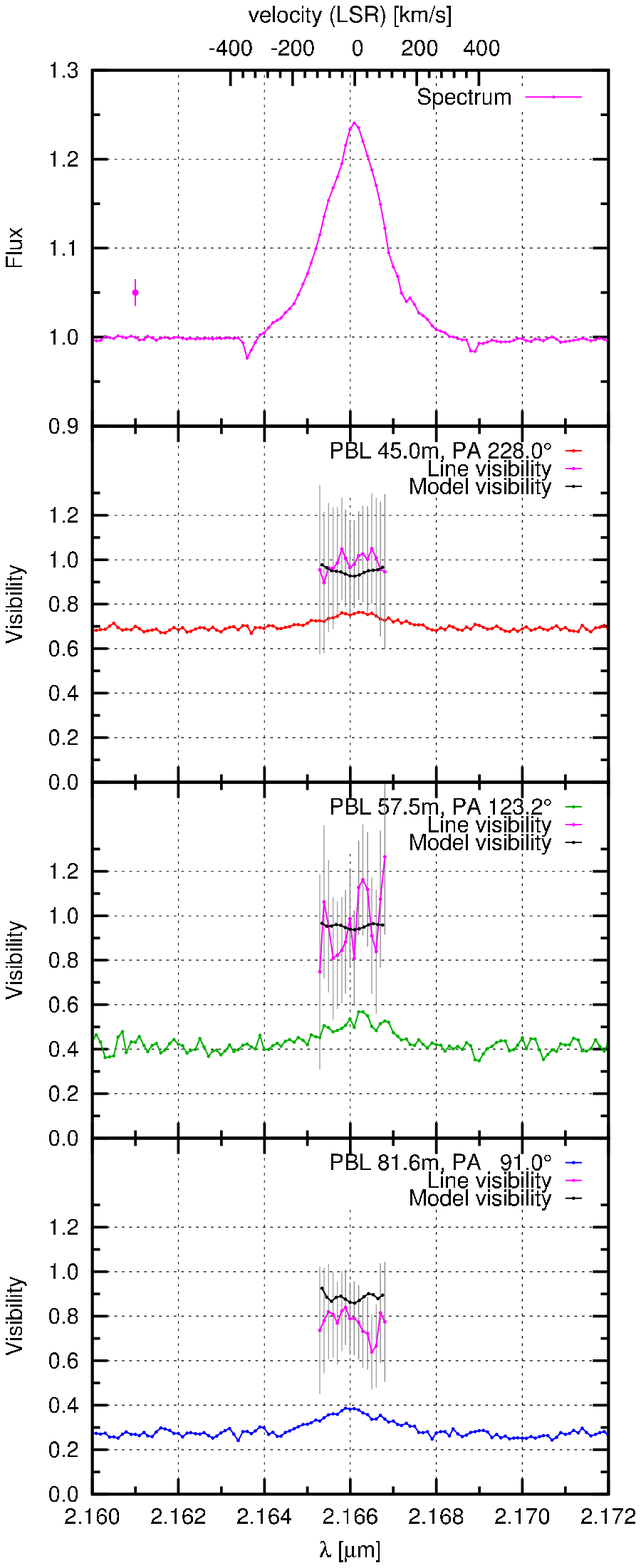}
        \includegraphics[width=0.85\columnwidth]{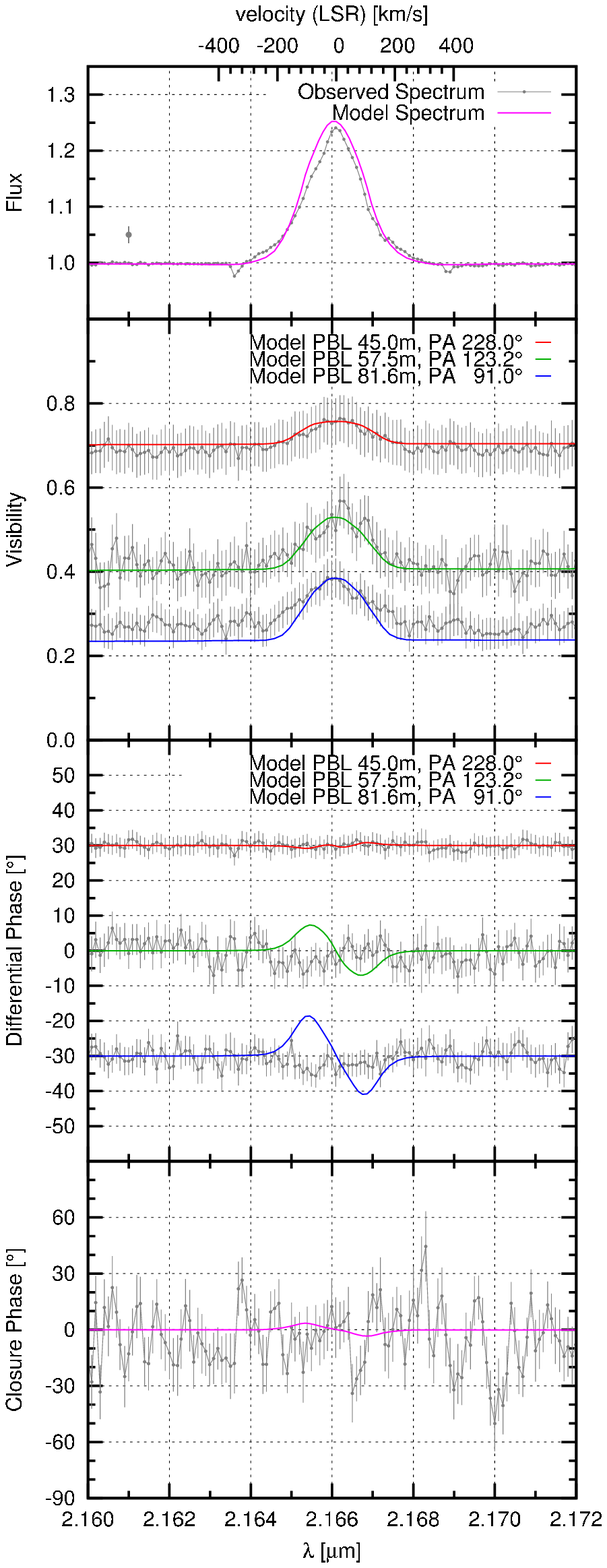}
        \caption{\label{fig:appendix_largeWN2}
                Same as Fig. \ref{fig:line_v} (left panel) and Fig. \ref{fig:best_fit} (right panel) but for model MW47 (see, Table\,\ref{tab:model_parameters}).
        }       
\end{figure*}

\section{Examples of computed hybrid models: disc wind plus magnetosphere}
\label{sect:appendix_hybrid_model}

\subsection{The magnestosphere}

A full description of the model employed here can be found in \cite{larisa14}. Here, only a brief description of the main model parameters is presented.

Our model considers a compact, disc-like rotating magnetosphere of height $h_m$ through which free-falling gas reaches the stellar surface at some altitude 
near the magnetic pole. The gas rotational velocity component ($u$) is described by
 \begin{equation}\label{ur}
    u(r) = U_0 (r/R_*)^p,
\end{equation}
where $U_0$ is the rotational velocity of the gas at the magnetic poles, $r$ is the distance from the star, and $p$ is a parameter. Finally, we assume a dependence of the electron temperature ($T_e$) of 
\begin{equation}\label{tm}
    T_e(r) = T_e(R_*) \exp(-r1),
\end{equation}
where $r1= ((r-R_*)/R_*)^q$,  $T_e(R_*)$ is the temperature of the gas near the stellar surface, and $q$ is a parameter.

\begin{table}[t]
\begin{minipage}[t]{\columnwidth}
\caption{\label{tab:magnetosphere_parameters} Magnetosphere model parameters}
\centering
\renewcommand{\footnoterule}{} 
\renewcommand*{\thefootnote}{\fnsymbol{footnote}}
\begin{tabular}{cc}
\hline \hline
Parameters                      &  MS6a             \\
\hline
$T_e(R_*)$\,(K)                 &  10\,000                  \\
$q$                             &  0.4                 \\
$U_*$\,(km/s)                   &  10                 \\
$U(r_T)/U_K$                    &  1                  \\
$r_c$\,(R$_*$)                  &  2.0 (0.02\,AU)     \\
$h_m$\,(R$_*$)                  &  1.5                \\           
$\dot{M}_{acc}$\,(M$_\odot$/yr) &  1$\times$10$^{-7}$  \\
\hline
\end{tabular}
\end{minipage}
\end{table}

\begin{figure*}
\centering
        \includegraphics[width=0.9\columnwidth]{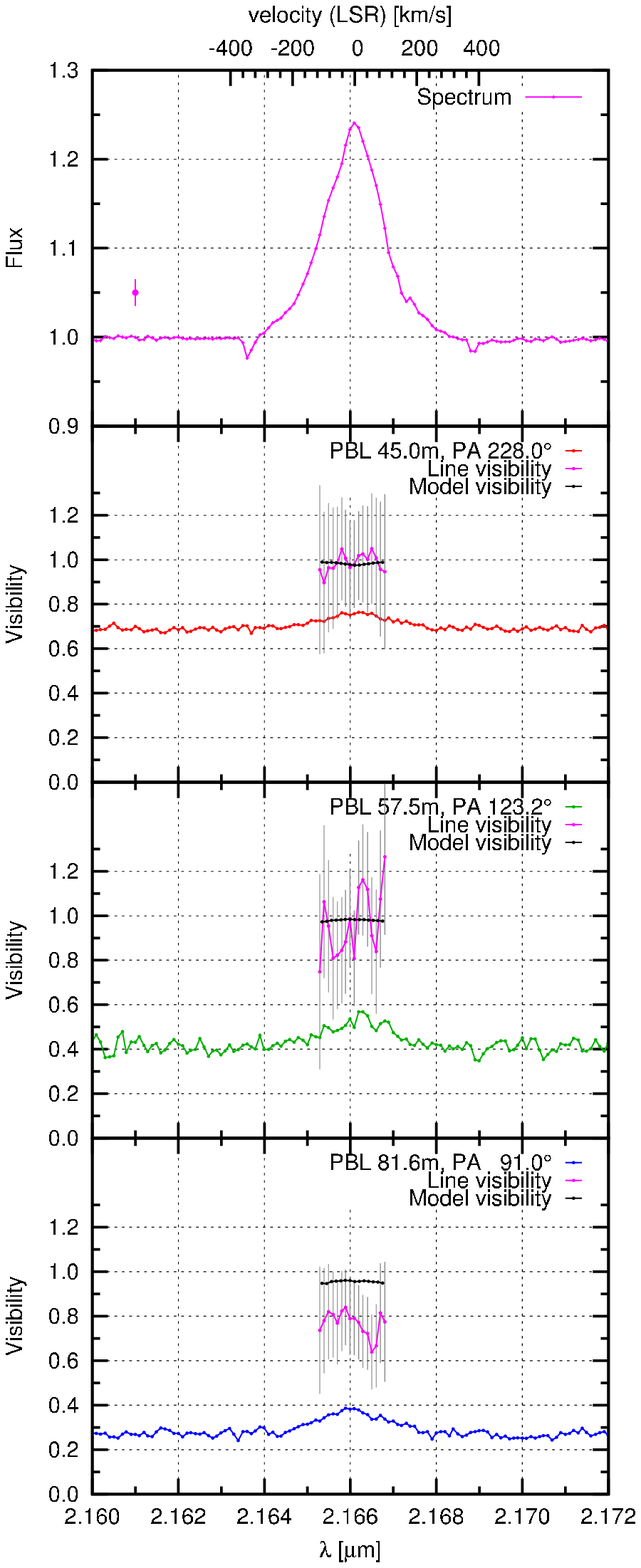}
        \includegraphics[width=0.85\columnwidth]{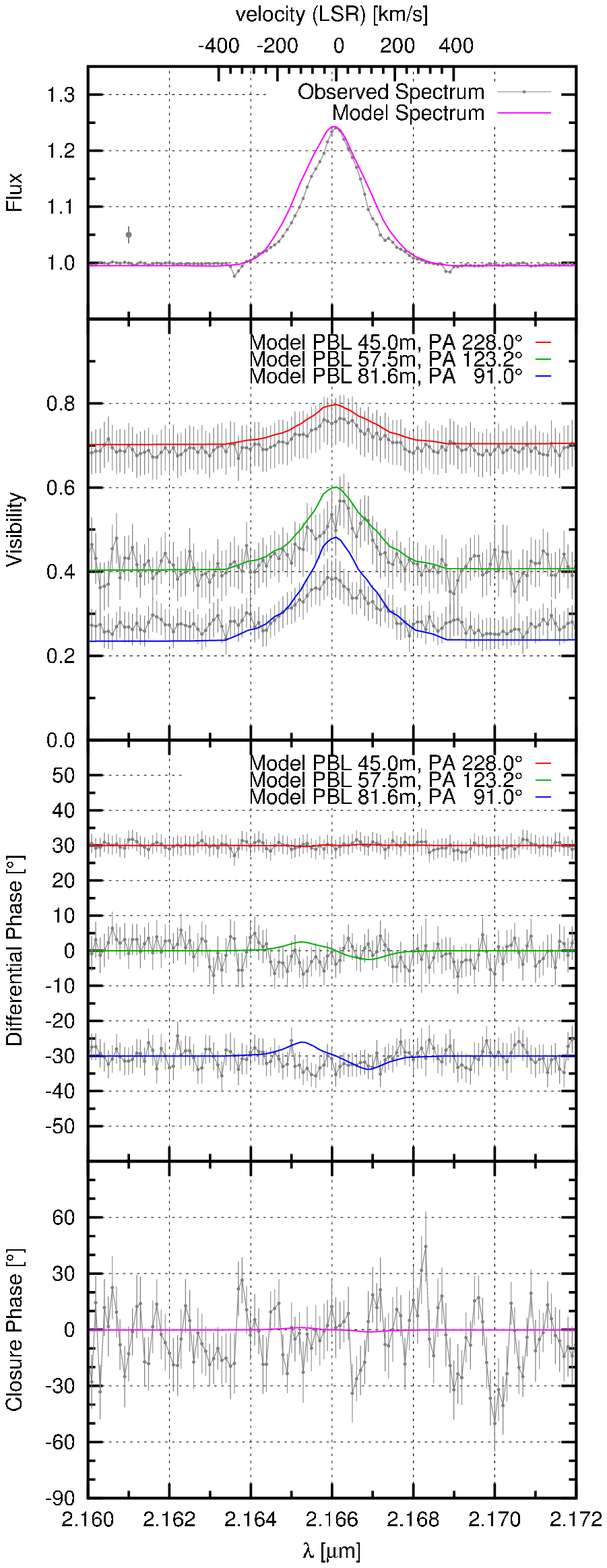}
        \caption{\label{fig:appendix_MW6+magnetosphere}
                Same as Fig. \ref{fig:line_v} (left panel) and Fig. \ref{fig:best_fit} (right panel) but for the hybrid model MW6+MS6a (see Tables\,\ref{tab:model_parameters} and 
                \ref{tab:magnetosphere_parameters}).
        }       
\end{figure*}

In our hybrid model, the magnetosphere is equivalent to a point source at our interferometer baselines. Thus, it accounts for the compact and unresolved \brg\ emission, whereas the disc wind 
component is responsible of the resolved \brg\ emission.

Because of the spread on the measured \macc\ values, and the large uncertainties in measuring this quantity ($\sim$20\%), an average value of 1$\times$10$^{-7}$\msyr\ was assumed in
our magnetosphere model.

\end{appendix}

\end{document}